\documentclass{ppai}

\usepackage{ppai}

\usepackage[utf8]{inputenc} % allow utf-8 input
\usepackage[T1]{fontenc}    % use 8-bit T1 fonts
\usepackage{hyperref}       % hyperlinks
\usepackage{url}            % simple URL typesetting
\usepackage{booktabs}       % professional-quality tables
\usepackage{amsfonts}       % blackboard math symbols
\usepackage{nicefrac}       % compact symbols for 1/2, etc.
\usepackage{microtype}      % microtypography
\usepackage{lipsum}
\usepackage{graphicx}
\graphicspath{ {./images/} }

\usepackage{inconsolata}

\usepackage{graphicx}
\usepackage{booktabs}
\usepackage{amsmath}
\usepackage{amssymb}
\usepackage{subcaption}
\usepackage{multirow}
\usepackage{multicol}
\usepackage{makecell}
\usepackage{bbm}\usepackage[normalem]{ulem}
\usepackage{enumitem}

\newcommand{\ours}{\emph{ObfuscaTune}}

\title{\emph{ObfuscaTune}: Obfuscated Offsite Finetuning and Inference of Proprietary LLMs on Private Datasets}

\author{
   Ahmed Frikha \\
   \And
   Nassim Walha \\
   \And
   Ricardo Mendes \\
   \And
   Krishna Kanth Nakka \\
   \And
   Xue Jiang \\
   \And 
   Xuebing Zhou\\
   \\
   Huawei Munich Research Center \\
\texttt{ahmed.frikha1@huawei.com}
}

\begin{document}
\maketitle
\begin{abstract}
 This work addresses the timely yet underexplored problem of performing inference and finetuning of a proprietary LLM owned by a model provider entity on the confidential/private data of another data owner entity, in a way that ensures the confidentiality of both the model and the data. Hereby, the finetuning is conducted offsite, i.e., on the computation infrastructure of a third-party cloud provider. We tackle this problem by proposing \emph{ObfuscaTune}, a novel, efficient and fully utility-preserving approach that combines a simple yet effective obfuscation technique with an efficient usage of confidential computing (only $~5\%$ of the model parameters are placed on TEE). We empirically demonstrate the effectiveness of \emph{ObfuscaTune} by validating it on GPT-2 models with different sizes on four NLP benchmark datasets. Finally, we compare to a naive version of our approach to highlight the necessity of using random matrices with low condition numbers in our approach to reduce errors induced by the obfuscation. 
\end{abstract}

% keywords can be removed

\section{Introduction}
Large Language Models (LLMs) such as {GPT-4} \cite{achiam2023gpt} are increasingly used due to their state-of-the-art performance in diverse tasks and productivity benefits \cite{noy2023experimental}. While LLMs excel in zero-shot and few-shot predictions with in-context learning \cite{mann2020language}, finetuning them on domain-specific data can significantly outperform foundation models in tasks like chip design~\cite{thakur2023benchmarking, wu2024chateda, liu2023chipnemo}.

Model providers keep their proprietary models private due to the exorbitant costs of training them\footnote{Training GPT-4 costed more than $\$100$M \cite{gpt4costs}}. To enable their users to customize or apply the proprietary models to their data, model owners provide finetuning and inference services, e.g., OpenAI finetuning API\footnote{https://platform.openai.com/docs/guides/fine-tuning} and GitHub Copilot\footnote{https://docs.github.com/en/copilot} respectively. Hereby, the users have to share their data with the model owners to use these services. Due to concerns of privacy leakage and competitive disadvantage, several users and commercial entities are not willing to share their private or confidential data. For instance, Samsung banned the usage of ChatGPT after sensitive code was leaked \cite{samsungleak}. Hence, approaches that enable the inference and finetuning of proprietary LLMs of one stakeholder on the confidential/private data of another stakeholder in a privacy-preserving way are crucially needed. 

Our contribution in the present work is threefold. First, we propose \emph{ObfuscaTune}, a novel and effective approach that combines a Trusted Execution Environment (TEE) with a simple yet effective obfuscation technique. Our proposed approach enables finetuning and inference of LLMs in a way that preserves the confidentiality of the model and the data with no utility loss and acceptable efficiency loss. {Second}, we empirically demonstrate the effectiveness of our method by validating it on GPT-2 models with different sizes on four NLP benchmark datasets. Hereby, only $5\%$ of the model parameters are placed on TEE. Although our experiments focus on the GPT-2 model family, ObfuscaTune can trivially be applied to any transformer-based LLMs and can potentially be generalized to further model architectures. {Finally}, we highlight the necessity of our \emph{optimized} obfuscation technique by comparing it to a naive obfuscation version.

\section{Problem statement and solution requirements}\label{problem}

The problem setting we consider involves three stakeholders: the model provider, the data owner and the cloud provider. The objective is to perform inference and finetuning of the proprietary LLM of the model provider on the confidential/private data of the data owner, in a way that ensures the confidentiality of both the model and the data. Due to the high computation and hardware costs required, we assume that the finetuning and/or inference is performed offsite, i.e., on the computational infrastructure of the cloud provider. In the considered threat model, we assume that the cloud provider is honest-but-curious, i.e., they will perform their task correctly but will try to find extra information about the other parties assets and data.

We define the following requirement that potential methods addressing this problem must fulfill: \begin{enumerate}[label=(\alph*)]
    \item Model confidentiality: prevent leakage of the proprietary model parameters
    \item Data confidentiality: prevent data leakage
    \item Utility: the performance and results of the inference and finetuning should be comparable with and without protection
    \item Efficiency: the computational time, memory footprint and communication should remain acceptable.
\end{enumerate}

\section{Related work} \label{related-work}
To the best of our knowledge, no prior work fulfills all of the requirements mentioned in Section \ref{problem} simultaneously. In the following, we discuss different categories of prior works. Prior approaches based on differential privacy (DP) for inference \cite{igamberdiev2023dp, majmudar2022differentially} and finetuning \cite{yu2021differentially} focus on protecting the data. However, they do not provide any protection for the model parameters and incur significant utility losses (Req. (a) and (c) are not fulfilled). Another line of work uses cryptographic techniques, e.g., multi-party computation (MPC) and homomorphic encryption (HE) \cite{li2022mpcformer, liu2023llms}. While the confidentiality of both the model and the data can be ensured, their substantial slowdown and communication costs are not suitable for real-time applications (Req. (d) is not fulfilled). Another proposal \cite{xiao2023offsite} considers sending a distilled version of the model to the client where adapter layers are finetuned on the confidential data. At inference time, the finetuned adapter are used in combination with the proprietary model on the server side. This approach does not protect inference data and leads to utility losses of up to $6\%$ (Req. (b) and (c) are not fulfilled). The closest approach to the present work combines Trusted Execution Environments (TEE) with a lightweight encryption to address federated learning settings \cite{huang2024fast}. However, such proposal protects only the finetuned LoRA parameters by using the TEE and deploys the proprietary LLM on the client-side fully or partially (Req. (a) is not fulfilled).

\section{Method}

\begin{figure*}[t]
    \centering
    \includegraphics[width=0.99\linewidth]{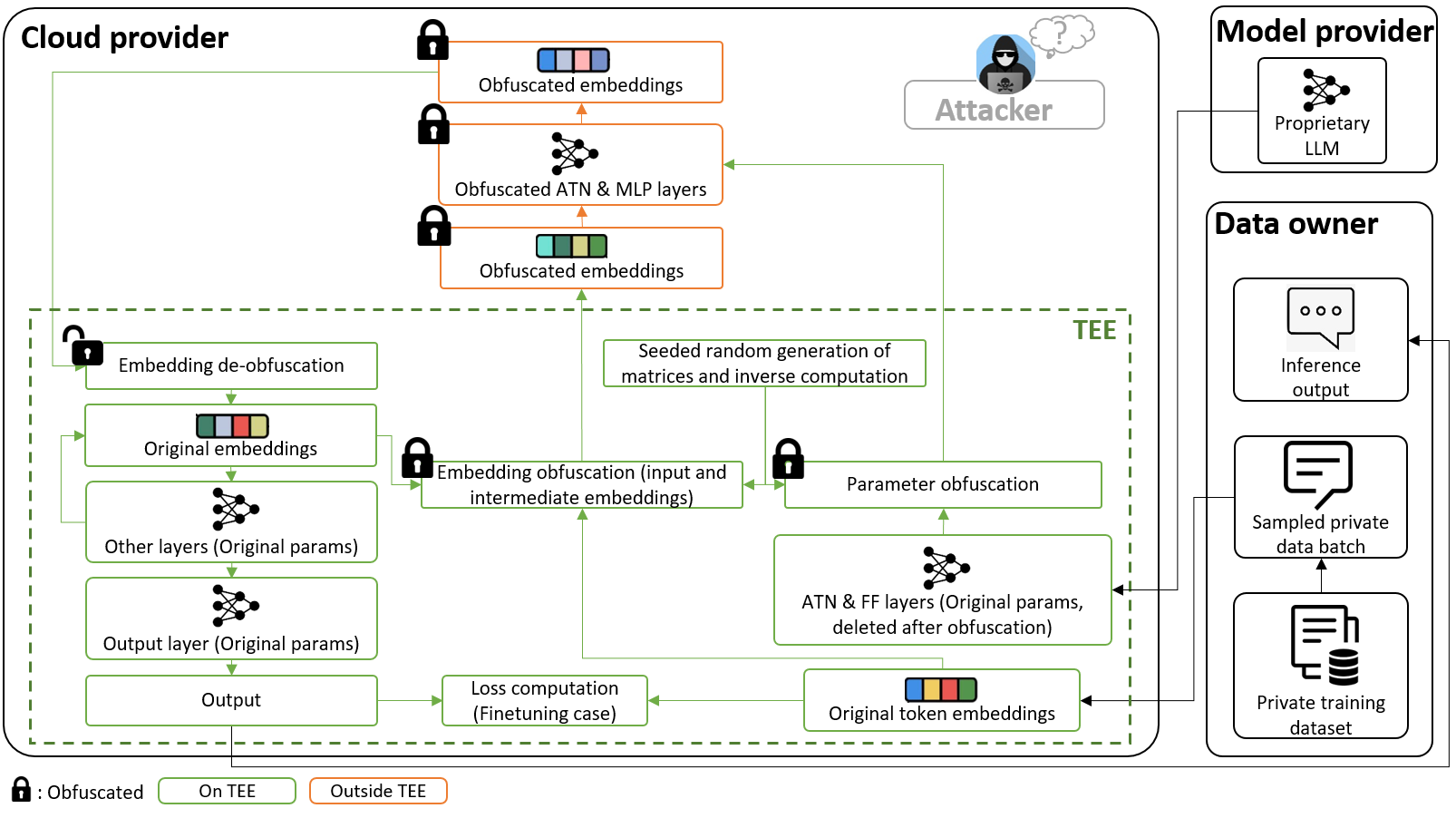}
    \caption{{\bf Overview of the proposed \emph{ObfuscaTune}}, composed by the three stakeholders: model provider, which seeks to keep the model confidential, data owner, which uses the model (finetuning or inference) while preserving privacy of their data, and cloud provider which provides the computation infrastructure, while potentially trying to eavesdrop on the data or steal the model. \emph{ObfuscaTune} provides the necessary protection by keeping very few components of the model within a TEE, and obfuscating the remaining ones, effectively and efficiently preventing data or model stealing.}
    \label{fig:highlevel}
\end{figure*}

To tackle the problem defined in Section \ref{problem}, we propose \emph{ObfuscaTune}, an approach that combines a Trusted Execution Environment (TEE) with a simple yet effective obfuscation technique, ensuring model and data confidentiality while preserving utility. Following prior works, we consider the TEE as an isolated secure zone on a potentially adversary host where the data, code and computation processes used are inaccessible from outside \cite{hou2021model, huang2024fast}. Note that modern high-end GPUs have TEE support \cite{h100}. Figure~\ref{fig:highlevel} presents an overview of the \emph{ObfuscaTune} approach, which we detail next.\\

\textbf{The model protection} is ensured as follows: the model provider sends the proprietary model to the TEE on the cloud provider infrastructure. Within the TEE, the highly parameterized attention and MLP layers are protected using our obfuscation technique that we detail later and then sent outside the TEE. Since large models do not fit inside the TEE, the model layers can be sent there batchwise to be protected before leaving it. The remaining low-parameterized layers, e.g., the input, output, normalization and dropout layers, are kept on the TEE. After these steps, all model parameters are protected, either by TEE or by the obfuscation, and the majority of model parameters are outside of the TEE. We note that the TEE is controlled by authentication that ensures that only the data owner can query the model. This prevents the cloud provider from querying the model to perform model stealing \cite{carlini2024stealing} or embedding inversion attacks \cite{li2023sentence, morris2023text}. \\

\textbf{The data protection} in \ours\ is conducted as follows: The data owner sends an encrypted batch of data directly to the TEE where it is first decrypted and then embedded using the model input layer. The resulting embedding is protected by our obfuscation method before leaving the TEE. The text tokenization can be conducted either before or after transmitting the data on the data owner side or in the TEE, respectively.\\

\textbf{The obfuscated feedforward pass} through one transformer block is executed as follows: Outside the TEE, the obfuscated data embedding is passed through the obfuscated model layers yielding an obfuscated intermediate embedding that is sent back to the TEE. The latter is then de-obfuscated and passed through the corresponding model layers on the TEE, depending on the model architecture. Subsequently, the resulting embedding is obfuscated again and leaves the TEE to be fed to the next transformer block. Finally, the output layer is applied in the TEE and the model output is sent back to the data owner (inference case) or used to computed the loss on the TEE and perform backpropagation and parameter updates (finetuning case).\\

\textbf{Our obfuscation method} obfuscates the model parameters and data embeddings by multiplying them with random matrices that minimize numerical errors. We begin by introducing the obfuscation method and later explain how we limit the numerical errors. Let's consider a multi-head attention layer and first focus on a single attention head with key, query, value layers parameterized by $W_k$, $W_q$ and $W_v$, respectively, and an embedding $X$ as its input. We obfuscate the embedding $X$ by multiplying it with a randomly generated matrix $R_a$, yielding $X^*$, and obfuscate the parameters $W_k$, $W_q$ and $W_v$ by multiplying them with the inverse of that random matrix, i.e., $R_a^{-1}$, yielding $W_k^*$, $W_q^*$ and $W_v^*$. Note that multiplying the obfuscated data embeddings $X^*$ with the obfuscated parameters, $W_k^*$, $W_q^*$ and $W_v^*$, leads to the same results, $Q$, $K$ and $V$, of the original non-obfuscated operations (Eq. 1-3). All obfuscation operations are applied inside the TEE. The remaining aforementioned operations are performed outside of the TEE.

\begin{figure*}[t]
    \centering
    \includegraphics[width=0.99\linewidth]{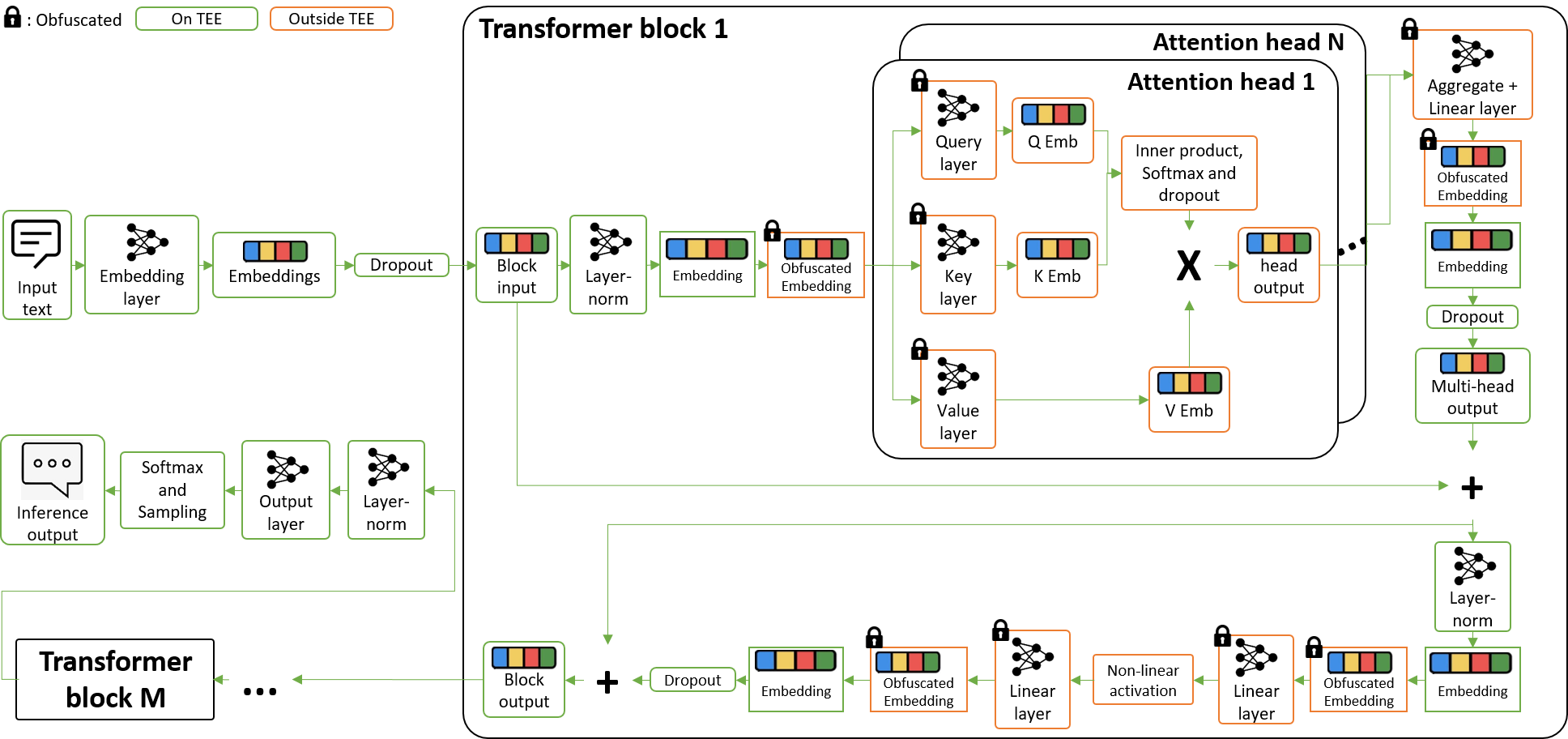}
    \caption{{\bf Detailed architecture of the GPT-2 with M layers using \emph{ObfuscaTune}}. Diagram blocks in green are within the TEE, while the orange are outside the TEE. This diagram illustrates how the data is successfully sent from and to the TEE, while being obfuscated while outside the TEE. Note that both the input text and output text are always within the TEE to prevent inversion attacks. Note that the non-linear activation applied after the first MLP (bottom) is applied on the de-obfuscated embedding. The same applies for the softmax non-linear function.}
    \label{fig:detailed}
\end{figure*}

The output $H$ of the attention head is computed (Eq. 4) and concatenated with the other heads outputs, yielding $H_{\mathrm{all heads}}$ \cite{vaswani2017attention}. $H_{\mathrm{all heads}}$ is then multiplied by the projection layer parameters $W_o^*$ that are obfuscated by another randomly generated random matrix $R_b$, yielding the obfuscated output $O^*$ of the multi-head attention layer (Eq. 5). Finally, this obfuscated output is sent to the TEE where it is de-obfuscated via multiplication with the inverse of the random matrix, i.e., $R_b^{-1}$. The bias term of this last projection layer has to be added after de-obfuscation and is therefore kept unobfuscated on the TEE. The obfuscation of the MLP layers of the proprietary LLM is conducted in an analogous manner to the obfuscation of the multi-head attention layers. Figure \ref{fig:detailed} shows an overview of all operations conducted in GPT-2 \cite{radford2019language} with annotations of which operations are performed inside or outside the TEE and on obfuscated or de-obfuscated variables. While we take GPT-2 as an exemplary LLM to showcase a concrete and detailed implementation of our approach, it is important to note that the application of ObfuscaTune to any transformer-based LLMs is straightforward, since recent LLMs share the same or similar building blocks. 

\begin{align}
Q  &= (X^T R_a) (R_a^{-1} W_q) = X^{*T} W_q^* \\
K  &= (X^T R_a) (R_a^{-1} W_k) = X^{*T} W_k^* \\
V  &= (X^T R_a) (R_a^{-1} W_v) = X^{*T} W_v^* \\
H  &= \mathrm{Dropout}(\mathrm{Softmax}(Q K^T)) V \\
O^*&= H_{\mathrm{all heads}}^T W_o^* \\
O  &= O^* R_b^{-1}
\end{align}

Note that using the same or different random matrices to obfuscate different transformer blocks does not impact our method. Also note that the layers that are kept on TEE involve non-linearities, e.g., layer-norm, and therefore cannot be applied to obfuscated variables since the subsequent de-obfuscation would not yield the same result. These layers have a low number of parameters compared to the attention and MLP layers placed outside of TEE, e.g., only ca. $5\%$ of the parameters of GPT2-XL are kept on TEE while $95\%$ are obfuscated and placed outside of TEE, in our experiments.\\

% Note that these embeddings cannot be inverted with state-of-the-art embedding inversion attacks \cite{li2023sentence, morris2023text} as these require a high number of model queries. This is not possible in this case, since querying the TEE requires authentication. 

\textbf{Security analysis:} Our security analysis is composed of three parts. First, typical privacy attacks such as membership inference~\cite{mireshghallah2022quantifying,zhang2024min} and embedding inversion attacks \cite{li2023sentence, morris2023text} are not applicable to our solution by design. Membership inference attacks are prevented by the authentication inside the TEE, i.e., only the data owner (who would be the victim of this attack) can query the model by authenticating themselves. SOTA embedding inversion attacks require a dataset of (text, embedding) pairs from the victim model to be able to reconstruct the text given the embedding. The collection of such a dataset by a potential attacker that would query the model with arbitrary data is also prevented by the TEE authentication mechanism. Note that the input text and output text of the data owner are always within the TEE to prevent such inversion attacks. We do not conduct empirical evaluations of these attacks, as this would not be possible without altering our proposed solution, e.g., removing authentication.

Second, it is not possible for the adversary to de-obfuscate the protected embeddings and model parameters. Note that all data embeddings and parameters that are accessible to the adversary, i.e., the ones that are processed outside of the TEE, are obfuscated, except for the intermediate embeddings $Q$, $K$ and $V$. A potential adversary would be interested in recovering a total of 5 unknown variables, i.e., the data embeddings $X$ and the model parameters $W_k$, $W_q$, $W_v$ and $W_o$, while having access to only 4 equations involving them (Eq. 1-3 and Eq. 5). Hence, it is not possible to compute them analytically. Moreover, it is not possible to come up with an approximate solution. The adversary would have to solve the system of equations $Q = X^TW_q$, $K = X^T W_k$, and $V = X^T W_v$, which are all non-linear and highly underdetermined since all of $X, W_q, W_k, W_v$ are unknown. Let $X, W_q, W_k, W_v$ be the real solution for the input embeddings and the weights respectively. Then for every invertible matrix $M$, $(M^{-1})^T X, M W_q, M W_k, M W_v$ would also be a solution of the system. This demonstrates that there is an infinity of possible solutions to that system of equations, and no analytic or numerical method can determine whether a solution corresponds to the real solution, i.e., the true weights/embeddings. The same applies to equation (5).

Third, we design our solution in a way that prevents any attacks that leverage frequency-based analysis, e.g., trying to deobfuscate the observed tokens by matching their frequencies with the frequency of the English language tokens. This is achieved by regularly performing model obfuscation with new randomly generated matrices, e.g., every hour or after a predetermined number of inferences. The model obfuscation can be performed very efficiently (less than 10 seconds on a middle range GPU for a GPT2-XL model), and can be made substantially faster by parallelizing the obfuscation of different layers.\\

\textbf{Minimizing numerical errors in the obfuscation:}. The proposed obfuscation can introduce numerical errors due to the multiplication with the random matrices and the inverse computation, which can propagate throughout the model and training iterations. Towards minimizing the error, we use only orthogonal random matrices, as they have the minimum condition number of $1$. The condition number $\kappa$ of a matrix $A$ is defined as $\kappa(A) = \frac{M}{m}$, where $M = \max_{x \in \mathbb{R}^n}{\frac{\|Ax\|}{\|x\|}}$ measures how much the mapping induced by that matrix can stretch vectors and $m = \min_{x \in \mathbb{R}^n}{\frac{\|Ax\|}{\|x\|}}$ measures how much it can shrink vectors. It determines how much a relative error in the input reflects on the output for solving linear systems, matrix inversion or matrix-vector multiplication \cite{golub2013matrix}. Such numerical errors get accumulated and increase with the number of sequential matrix multiplication operations, i.e., the deeper the model the higher the accumulated error. We minimize the numerical errors by minimizing the condition number of the random matrix. In this work, we consider the condition number w.r.t.\ the $\ell_2$ norm. Since orthogonal matrices induce isometries, i.e $\|Ax\|_2 = \|x\|_2$ for all $x$, we get $\kappa(A) = 1$ for every orthogonal matrix $A$. Note that singular matrices have the highest (worst) possible condition number, which is $\infty$, since for a singular matrix $A$,  $m = \min{\frac{\|Ax\|}{\|x\|}} = 0$. On the other side, from the definition we see that the lowest possible $\kappa$ is 1. 

We ensure that our random matrices $R_a$ and $R_b$ have a condition number of $1$ by setting them to be the $Q$ matrix computed by applying a $QR$-decomposition to a randomly generated matrix, as $Q$ is always orthogonal. In this case, the inverse computation is fully error-free since the inverse of an orthogonal matrix is its transposed version which is an error-free operation. \\

\section{Experimental evaluation and discussion}\label{exps}

The conducted experiments aim to address the following key questions: $\textbf{(a)}$ What is the impact of applying \emph{ObfuscaTune} on utility, i.e., how do models finetuned with \emph{ObfuscaTune} compare to the normally finetuned models? $\textbf{(b)}$ How does our obfuscation method using orthogonal random matrices compare to naively using any random matrices? 

We apply our method to GPT2~\cite{radford2019language} models with different sizes, ranging from 117 million to 1.5 billion parameters. We implement \emph{ObfuscaTune} on top of the nanoGPT implementation \cite{nanoGPT}. All our experiments perform LoRA-finetuning \cite{hu2022lora}. Hereby, the LoRA parameters are randomly initialized and placed outside of the TEE. We apply LoRA to all linear and attention layers. Further hyperparameters are specified in the appendix. 

In each \emph{ObfuscaTune} experiment, we use 2 GPU devices, one that is placed outside of TEE and another that simulates the TEE. We believe this is reasonable since high-end GPUs have TEE support \cite{h100}. We evaluate the finetuning with \emph{ObfuscaTune} and with a naive version that uses any random matrices on 4 question-answering benchmark datasets, including WebQuestions (WebQs) \cite{berant2013semantic}, OpenBookQA (OBQA) \cite{mihaylov2018can}, PIQA \cite{bisk2020piqa} and SciQ \cite{welbl2017crowdsourcing}. 
We evaluate all models using \texttt{lm-eval-harness}\footnote{https://github.com/EleutherAI/lm-evaluation-harness}. 

\begin{table}[ht]
\centering
\small
\begin{tabular}{lcccc}
\toprule
{\bf Setting} & {\bf WebQs} & {\bf OBQA} & {\bf PIQA} & {\bf SciQ} \\
\hline
\multicolumn{5}{c}{\bf GPT2-Small}\\
Unprotected & 16.0 & 23.0 & 64.1 & 91.1 \\
Protected (random) & 0.0 & 15.4 & 53.1 & 19.7\\
Protected ({\bf ours}) & 16.8 & 23.6 & 64.8 & 91.7\\
\hline
\multicolumn{5}{c}{\bf GPT2-Medium}\\
Unprotected & 24.1 & 29.2 & 69.1 & 92.2 \\
Protected (random) & 0.0 & 14.4 & 52.0 & 20.0\\
Protected ({\bf ours}) & 24.5 & 28.6 & 68.9 & 92.4 \\
\hline
\multicolumn{5}{c}{\bf GPT2-Large}\\
Unprotected & 30.0 & 35.0 & 72.1 & 93.3 \\
Protected (random) & 0.0 & 14.4 & 52.0 & 19.7 \\
Protected ({\bf ours}) & 29.7 & 32.2 & 72.3 & 93.0 \\
\hline
 \multicolumn{5}{c}{\bf GPT2-XL}\\
Unprotected & 32.4 & 34.2 & 74.1 & 93.5 \\
Offsite Tuning \cite{xiao2023offsite} & 19.9 & 28.2 & 73.6 & 93.2 \\
Protected (random) & 0.0 & 14.8 & 52.5 & 20.5 \\
Protected ({\bf ours}) & 32.6 & 33.2 & 73.9 & 93.6 \\
\bottomrule
\\
\end{tabular}
\caption{Test accuracy results (\%) yielded by normally finetuned models (unprotected) and models which are protected by \emph{ObfuscaTune} as well as a naive version of our method that uses an arbitrary random matrix with a non-optimized condition number (random).}
\label{tab:results}
\end{table}

Table \ref{tab:results} presents our main experimental results. We find that models finetuned with our method achieve a performance comparable to models finetuned without model and data protection. This observation is consistent across all model sizes and benchmark datasets. Besides, models that are finetuned with a naive method that uses arbitrary random matrices incur substantial utility loss due to the high accumulation of errors. For GPT2-XL, we additionally compare to Offsite Tuning \cite{xiao2023offsite} and find that ObfuscaTune outperforms it significantly. It is worth noting that, besides the utility loss, Offsite Tuning does not protect inference data and discloses a distilled version of the proprietary model. As discussed in Section \ref{related-work}, to the best of our knowledge, all prior approaches infringe at least one of the requirements of a viable solution to the addressed problem (Section \ref{problem}).

\begin{table}[ht]
\centering
\small
\begin{tabular}{ccccccc}
\toprule
{ CN} & 1 & 8 & 32 & 128 & 160 & random \\
\hline
{ Accuracy} & 16.8 & 15.5 & 15.2 & 14.7 & 0.3 & 0.0\\
\bottomrule
\\
\end{tabular}
\caption{Test accuracy results (\%) yielded by GPT2-small models finetuned on WebQs with \emph{ObfuscaTune} using matrices with different condition numbers (CN).}
\label{tab:condition-number}
\end{table}

\textbf{Impact of different condition numbers:} We investigate the impact of using different obfuscation matrices with different condition numbers and present our results in Table \ref{tab:condition-number}. For this, we generate different random matrices with different condition numbers that we pre-define. We provide details on this in Appendix \ref{sec:condition}. Our findings empirically confirm that higher condition numbers deteriorate performance due to larger numerical errors.

\textbf{Applicability to larger models:} We measure the percentage of model parameters present on TEE after model obfuscation to be $\boldsymbol{5.2\%}$ for GPT2-XL, highlighting a substantial efficiency increase compared to naively shielding the entire model inside the TEE. This efficiency increase becomes even larger for models with a higher number of parameters, since the proportion of the number of obfuscated parameters placed outside of the TEE w.r.t. the total number of parameters becomes higher. For instance, applying ObfuscaTune to the LLama 3 405B model \cite{dubey2024llama} would result in less than $0.5\%$ of the parameters being placed on TEE. It is important to note that placing less parameters on GPUs with TEE support leads to cost savings for the infrastructure provider, since other cheaper GPUs can be used for the obfuscated parameters. Our empirical results show that model size does not affect method accuracy, as evidenced in Table 1. Regarding the model architecture, we believe that ObfuscaTune is applicable to any transformer-based LLMs (incl. encoder-decoder and encoder-only models). This is due to the fact that most recent LLMs, e.g., Llama 3, are composed of the same building blocks. In addition to being LLM-architecture-agnostic, ObfuscaTune can potentially be generalized to further model architectures beyond transformer-based models.

Finally, we measure the runtime of the finetuning and find that using \emph{ObfuscaTune} leads to a slowdown of $1.5$x to $4.3$x, for GPT2-small and GPT2-XL respectively. While the slowdown incurred by ObfuscaTune grows with an increasing number of layers, it is worth noting that the number of layers in recent LLMs does not increase linearly with the number of parameters. For instance, the Llama 3 models with 8B, 70B and 405B parameters have 32, 80 and 126 layers, respectively. It is important to also note that the slowdown incurred by ObfuscaTune is substantially lower than slowdowns yielded by cryptographic techniques, e.g., ca. $10^2$ using MPC \cite{knott2021crypten} and $10^5$ using HE \cite{lou2021hemet} with models that are significantly smaller (some orders of magnitude smaller).

\section{Conclusion}
This work tackled the timely but underexplored problem of performing offsite inference and finetuning of a proprietary LLM owned by a model provider entity on the confidential/private data of another data owner entity, in a way that ensures the confidentiality of both the model and the data. Our proposed approach, ObfuscaTune, achieves this by combining a simple yet effective obfuscation technique with efficient use of confidential computing (only $~5\%$ of the model parameters are placed on TEE). Our extensive empirical evaluation on four NLP benchmark datasets and different models highlights the effectiveness of our method and emphasizes the importance of using random matrices with low condition numbers to preserve high utility. In future work, we will investigate the effectiveness of our approach to RAG-systems.

\bibliographystyle{unsrt}  
\bibliography{references}  %%% Remove comment to use the external .bib file (using bibtex).



% George Kour and Raid Saabne.
% \newblock Real-time segmentation of on-line handwritten arabic script.
% \newblock In {\em Frontiers in Handwriting Recognition (ICFHR), 2014 14th
%   International Conference on}, pages 417--422. IEEE, 2014.

% George Kour and Raid Saabne.
% \newblock Fast classification of handwritten on-line arabic characters.
% \newblock In {\em Soft Computing and Pattern Recognition (SoCPaR), 2014 6th
%   International Conference of}, pages 312--318. IEEE, 2014.

% \bibitem{hadash2018estimate}
% Guy Hadash, Einat Kermany, Boaz Carmeli, Ofer Lavi, George Kour, and Alon
% \newblock Estimate and replace: A novel approach to integrating deep neural
%   networks with existing applications.
% \newblock {\em arXiv preprint arXiv:1804.09028}, 2018.

% \end{thebibliography}

\appendix
\label{sec:appendix}

\section{Hyperparameters}
We train all models for 10 epochs. We perform validation at the end of every epoch and use early stopping with a patience of 3. We use a learning rate of ${3e-5}$ and a batch size of 1. We keep the other hyperparameters unchanged from \cite{nanoGPT}. For LoRA, we use the hyperparameters: $r=16$, $\alpha=32$ and apply dropout with $0.05$. We did not perform hyperparameter tuning, which highlights the robustness of our method. We did all experiments on middle-range GPUs. Each experiment took between 1 and 8 GPU hours, depending on the model size and dataset.\\

\section{Random matrix generation with a pre-defined condition number}\label{sec:condition}

Let $\sigma_{max}(A)$ and $\sigma_{min}(A)$, respectively, be the largest and the smallest singular values of the matrix $A$. For the $\ell_2$-induced operator norm norm the following holds :
\begin{equation*}    
\|A\|=\max{\frac{\|Ax\|}{\|x\|}} = \sigma_{max}(A).
\end{equation*}
On the other hand, for A square and non-singular
\begin{align*}
    \min_{x} \frac{\|Ax\|}{\|x\|}
    &= \min_{y} \frac{\|y\|}{\|A^{-1} y\|} \\
    &= \frac{1}{\max_{y} \frac{\|A^{-1} y\|}{\|y\|}} \\
    &= \frac{1}{\|A^{-1}\|} \\
    &= \frac{1}{\sigma_{max}(A^{-1})} = \sigma_{min}(A).
\end{align*}
 Finally we get for every square and non-singular matrix $A$:
 \begin{equation*}
     \kappa(A) = \frac{\sigma_{max}(A)}{\sigma_{min}(A)}
 \end{equation*}

\textbf{To generate random matrices $R$ of a given predefined condition number $\kappa(R)$}, we leverage this last equation as follows. First, we generate random matrices $A$ and $B$ using the standard normal distribution. We then apply QR-decomposition on $A$ and $B$ to generate two orthogonal matrices $Q_A$ and $Q_B$. We then choose a random positive value for the largest singular value of the final matrix $R$ and set $\sigma_{min}(R) = \frac{\sigma_{max}(R)}{\kappa(R)}$. The remaining singular values can be sampled randomly from the uniform distribution between $\sigma_{min}(R)$ and $\sigma_{max}(R)$. Then we construct the diagonal matrix $S$ with the singular values on the diagonal. Note that $S^{-1}$ is the diagonal matrix with the inverses of the singular values on the diagonal. Then we define $R$ to be having the following singular value decomposition:
\begin{equation}
    R = Q_A S Q_B.
\end{equation}
And can calculate $R^{-1} = Q_B^T S^{-1} Q_A^T$ with minimal rounding errors. We use this approach to generate random matrices of a given condition number and monitor the effect of the condition number on the test accuracy of the final model. The results presented in table \ref{tab:condition-number} show indeed that it is crucial to have a low condition number, otherwise the training degenerates.

\end{document}